# Search for Hidden Knowledge in Collective Intelligence dealing Indeterminacy Ontology of Folksonomy with Linguistic Pragmatics and Quantum Logic

Massimiliano Dal Mas

*me @ maxdalmas.com*

**Abstract**

Information retrieval is not only the most frequent application executed on the Web but it is also the base of different types of applications. Considering collective intelligence of groups of individuals as a framework for evaluating and incorporating new experiences and information we often cannot retrieve such knowledge being tacit. Tacit knowledge underlies many competitive capabilities and it is hard to articulate on discrete ontology structure. It is unstructured or unorganized, and therefore remains hidden.
Developing generic solutions that can find the hidden knowledge is extremely complex. Moreover this will be a great challenge for the developers of semantic technologies.
This work aims to explore ways to make explicit and available the tacit knowledge hidden in the collective intelligence of a collaborative environment within organizations. The environment was defined by folksonomies supported by a faceted semantic search. Vector space model which incorporates an analogy with the mathematical apparatus of quantum theory is adopted for the representation and manipulation of the meaning of folksonomy. Vector space retrieval has been proven efficiency when there isn't a data behavioural because it bears ranking algorithms involving a small number of types of elements and few operations.
A solution to find what the user has in mind when posing a query could be based on "joint meaning" understood as a joint construal of the creator of the contents and the reader of the contents. The joint meaning was proposed to deal with vagueness on ontology of folksonomy indeterminacy, incompleteness and inconsistencies on collective intelligence.
A proof-of concept prototype was built for collaborative environment as evolution of the actual social networks (like Facebook, LinkedIn,..) using the information visualization on a RIA application with Semantic Web techniques and technologies.

## 1. Introduction

Web Knowledge is actually recognized as the fourth production factor for the global economy, managing knowledge is as important as the traditional management of labour, capital and materials. A prerequisite is that knowledge should be computer-accessible. This coincides with the vision of the Semantic Web (SW) at large that can make an important contribution to knowledge management.

A "collective intelligence" could be defined as the ability of a group to solve problems than its individual members cannot.
In collective intelligence when we do knowledge work on the web we should want to discover other people doing the same work, perhaps to share or connect up.

This was the vision that launched the Web, and it drives the goal of accelerating human knowledge and understanding. This use case requires that there be a common conceptualization of what tagging an object means and at least some way for a service to correlate or connect tag data from one application to another.
It should be hard to have a single, standardized way to collect, interpret, or use tag data. But we can build the substrate for an ecosystem of tagging that will lets us innovate and work toward the vision of an open tagosphere. Tagging is about identifying and formalizing a conceptualization of the activity, and building technology that commits to the ontology at the semantic level.
From the user's point of view, tagging is an activity in which you label some content you create or experience with one or more labels, or tags.
So to clarify the meaning of tagging, we would design a different sort of relation or family for "meta tagging" or whatever it might be called. One system might use a tag-on-tag notion to mean "this tag is a synonym of that tag" and another system might have a notion of "this tag represents a cluster of other tags". There is no requirement that all systems share the same notions; a successful

———

If you have read this paper and wish to be included in a mailing list (or other means of communication) that I maintain on the subject, than send e-mail to: *me @ maxdalmas.com* knowledge sharing agreement only requires that they clearly identify the differences when they share data.
*Ontologies* are as much about reasoning about incompatibilities as about finding commonalities. Ontologies are conceptual specifications that enable multiple, independently developed databases of carefully categorized artefacts to interoperate, and for agents to reason about the differences among the vocabulary used.
Ontology is an attack on top down categorization as a way of finding and organizing information. For the task of finding information, taxonomies are too rigid and purely text-based search is too weak. Tags introduce distributed human intelligence into the system. As others have pointed out, Google's revolution in search quality began when it incorporated a measure of "popular" acclaim -- the hyperlink -- as evidence that a page ought to be associated with a query. When the early webmasters were manually creating directories of interesting sites relevant to their interests, they were implicitly "voting with their links." Today, as the adopters of tagging systems enthusiastically label their books and photos, they are implicitly voting with their tags. This is, indeed, "radical" in the political sense, and clearly a source of power to exploit.

The meaning of a linguistic expression in a language is determined by how it is used by a community of competent rational speakers of the language. On the holistic view the meaning of a statement involves a complicated network of connections to other statements with different components and beliefs having to considered the behavioural circumstances.
Ontologies can capture the semantics of a set of terms used by some community: but meanings change over time, and in any case it will not be easy to negotiate semantic definitions that can be accepted by an articulated community.
Human meanings are based on individual experiences, and logical axioms can only partially reflect them.
*Folksonomy* is an attack on bottom up categorization. The praise for folksonomy is really the observation that we now have an entirely new source of data for finding and organizing information: user participation.
Semantic Web technologies can improve the effectiveness of collaborative environments within corporate intranets addressing the folksonomy indeterminacy respect user (*reader*) ontology on the Semantic Search, having a defined environment meanings shared by the relevant communities, and reasonably stable over significant periods of time. This work investigates a possible solution in a defined environment, like a community, to find principle to be applied in the open Web.

Online archives are characterized primarily by their content being distributed across hundreds of single websites; among other things, that makes the power of orientation difficult. The *Semantic Map* can orders all the information on the platforms and shows information as a mesh of semantic relationships through performative interfaces.
Semantic Web databases are not (completely) disorderly and there are many ways to optimize the search for matching triples to a graph pattern. Quantum logic could solved problems related to Grover's Algorithm [1] speeding up queries of disorderly databases. The larger the triple store, the more compelling the case for using some kind of quantum search algorithm to find matches in conjunction with inference engines.

Among problem that arises it is necessary to define an indexing schemes to structure the triples stored and how to deal with "quantum superposition", a fundamental law of quantum mechanics that defines the collection of all possible states that an object can have. Furthermore any returned conclusion, as process of evolution/action that "collapse" with the choice of one value excluding other, should be relative to the axioms of the referred ontology.
*Ontologies* are a critical aspect of the Semantic, being a formal models of human domain knowledge they are difficult to build.
Often there is no a discrete structure for single correct mapping on human knowledge. An ontology is a formal specification of a conceptualization [2] that can be understood as an abstract representation of the world or domain we want to model for a certain purpose. Ontology-design tasks follows some rules but those cannot be comprehensive of the *tacit knowledge*.
When we need such knowledge, we often cannot retrieve it because, being tacit, it is unstructured or unorganized, and therefore remains hidden. In order to understand the full potential of tacit knowledge, we can consider the difficulties when a key person leaves a company and takes important knowledge assets with him (or her.)

Developing generic solutions that can find the hidden knowledge is extremely complex but this will be the biggest challenge for the developers of semantic technologies.
This work wants to propose a direction for solutions able to make explicit and available the tacit knowledge hidden in collaborative environment within organizations using the folksonomy categorization from the user participation.
These examples prove how much can be saved, in terms of time and costs, by an application able to find the tacit knowledge by a group, organize the contents, and make them accessible and usable in the future.
Semantic Web can have a key role but under present conditions it requires considerable customization and tuning, an investment that can afford only big companies. Semantic Web technology can be used for knowledge management on a company wide scale with company intranets.
Rather than being limited to company-wide intranets Semantic Web contains the promise that the knowledge management technologies would also be used towards a worldwide Semantic Web.
Much of the same technology that was developed for the purposes of company-wide knowledge management has also been shown to be useful on a much larger scale. Also small and medium businesses could benefit from tacit knowledge that represents a really relevant cost. As alternative, tools should support ontology evolution and iteration. The proposed interaction paradigm for semantic search aims to deal with vagueness, incompleteness and inconsistencies with query interpretation based on the pragmatic of dialogue.

In *social networks*, known as *Web 2.0*, new knowledge is created by sharing ideas as information, that is done by data. While in the Semantic Web data is shared between different social networks. Using the Semantic Web for the social network can help even small and medium companies in sharing information and knowledge creating communities of people with similar interests.
A predefined structure to organize information is not applicable for a social network that can be represented as a graph in evolution where users can create different kinds of classification on the fly from the previously decided. It is so necessary to assign an item to multiple parameters each representing an aspect or a “facet” of the information.
In such a scenario a new springtime arose on the concepts of "facets" and "faceted classification" as multidimensional classification developed by Ranganathan in the context of classical librarianship.
According to faceted classification the user determines his/her own aggregation of various parameters. The semantic relations between facets may express the user choices by a relationship of facets: “I'm looking at a facet which has certain relations with other facets”. That can be expressed in RDF by subjects, predicates and objects. So information should be semantic allowing the user to give value to what for him/her is important.
A faceted system brings us to the subject of scale. In information retrieval, size matters, and we only learned recently how much.

In any knowledge management initiative, technology alone is not sufficient but it is necessary a methodology that should be used in order to effectively apply the technology to exploit the Semantic Web. For knowledge management it is necessary to have different kind of tools: an ontology editor to capture human expertise, an environment for collaborative knowledge sharing, tools for automated concept extraction from knowledge sources, support for semantic search and navigation through knowledge sources, among others. The representation languages should be expressive, logically well-founded, and compatible with current IT standards such as XML.

Much exciting work remains to be done to make the transition from the theories into real practice, in real industry, on the real web. The tools are in the main prototypes, the methodology needs to mature, the case studies have been relatively costly, labour intensive and have required input from skilled specialists.
This paper wants to integrate folksonomies and faceted navigation having to balance the rigidity of facets built from a controlled vocabulary with the potential anarchy of raw folksonomies. To develop this proposal it is necessary to improve the quality of the individual tags on folksonomies given a classification order according different facets. This work considers a mechanism based on the “joint meaning”, as commitment providing a way to connect speech acts from the speaker and the reader so from who contribute to the web, called as *speakers*, and who search the web, called as *reader*. The system will allow dynamic selection of categories using better an auto-classification developed integrating the RIA features of the Web 2.0 with meaning of the Semantic Web (Web 3.0).
*Joint meaning* will be used as procedure to deal with vagueness on indeterminacy given the ability to extract semantic faceted metadata and create semantic associations leading to better search, integration and analysis.
In this paper was reviewed the recent developments in applying the geometric and quantum methods as collaborative interaction paradigm dealing indeterminacy in Semantic Search.

In the joint meaning the semantic relation was determined between a *reader {u}* and a speaker, of the *speakers U–{u}* group, that define his/her lattice by a triple composed by facets, tags and the relative incidence relations of context ($t_n$, $f_n$, $i_n$) that was determined by the semantic search. The speakers lattice was chosen according to the joint meeting between all the speakers lattice. These were settled as a framework in a Dilberth space to determine their compatibility and incompatibility. Inside the speakers lattice chosen is used the incidence relation of context.
Multiple matching was then disambiguated by updating a similarity degree associated to the incidence relation.

The contributions of this paper are as follows:

- it introduces the tacit knowledge hidden in the collective intelligence
- it shows the approach proposed for tacit knowledge with the relation between facets and folksonomies
- it introduces the use of the joint meaning concept on the topic composed by incidence of concepts of every speaker
- it discusses some quantum mechanics phenomena presenting the analogies between the information retrieval with facets and folksonomies
- it describes the architecture and the implementation of the developed prototype
- finally, it draws conclusion and suggest directions for future works

## 2. Facets and folksonomy

This section introduces the notation and the necessary background for this article. It first introduces the faceted classification and then the folksonomy.
The faceted classification of an object by exploiting a system of attributes (metadata) representing each one aspect or property is able to describe exhaustively the object itself.
Faceted classification is a method of classification of the distinctive value of which consists in being open and adaptive.
The main distinguishing features of a faceted system are summarized as:

- *multidimensionality*: inverse to traditional systems, in faceted systems, each object is classified according to a plurality of attributes called facets, while not knowing the name or the location of an object, it is nevertheless possible to understand and achieve it, describing it through a set of categories (facets) mutually exclusive

- *persistence*: these attributes / facets are essential properties of the object and persistent, so the impact (on the classification scheme) of any change (within the classification, workflow, etc..) is strongly reduced to zero being possible to build persistent relationships between the different facets, such as to provide a knowledge representation in the system: eg. `John Dale` (facet PERSON) `<is an employee of the>` Company Z (facet COMPANY) and `<develop>` Applications XML (facet PRODUCTS).
- *scalar*: it is always possible to add a new descriptive facet of a new aspect of the object to the already established at an earlier stage (open system)
- *flexibility*: there is a plurality of parallel access keys (*facets*), each object can be found using a single search attribute (or *facet*) at a time, or a combination of multiple attributes been not necessary to know the name / class and where the object is placed (in a context of rapid change this is an advantage)
- *not hierarchical sort order* (from general to particular, from the inside of an object part, etc.) to prevent limit search results and describe relations between objects (knowledge representation in the system)

Considering a faceted classification only as a theoretical apparatus coined by science books is limitative. This approach, in fact, is the formalization of a technique of communication that we often use in a wide range of contexts, from the organization of personal information.

The faceted classification has important advantages over other systems in particular: *multidimensionality*, *persistence*, *flexibility* and *scale*. These features prevent the deterioration of repository avoiding that changes have negative repercussions on the information organization.

Facets are orthogonal with mutually exclusive dimensions (for example: a seminar is not a person, is not a document, is not a place) using an active interface with a dynamic combination of search and browse applied at search time with a post-coordination and not a pre-coordination as in the Advanced Search. An Advanced Search could be an important addition to Search / Browse but requires adding lots of metadata and to understand users for the information architecture.

Faceted interfaces result more intuitive for its simplicity of internal organization that allows multiple perspectives with the ability to handle compound subjects.

An internal facet structure reflects current usage as user of a community, so they can understand different structures matching the structure to domain and task.

Having a precision of unit values, it allows flexibility for additions of new subjects, facets, entities at any point in the system with different kinds of categorization: chronological, alphabetical, spatial, simple to complex, size or quantity, hierarchical, canonical. The implementation of faceted user interface usually have some disadvantages: loss of browse context, difficult to grasp scope and relationships, difficulty of expressing complex relationships, limited domain applicability as type and size.

Ontologies and folksonomies are two separate approaches to two different types of problem, although some of the functionality of ontologies can be taken over by folksonomies in a number of contexts. [3, 4, 5]

Folksonomies are a variant on the keyword search theme, and an attempt at information retrieval but till now they cannot be used to retrieve documents relevant to the concept in which I am interested.

Ontologies are attempts to regulate parts of the world of data, and to allow mappings and interactions between data held in disparate formats or locations, or which has been collected by different organisations under different assumptions.

Furthermore there is a perception that folksonomies evolve organically and painlessly whereas ontologies are high maintenance and high overhead (with relative costs).

The sets of problems they are approaches may overlap.

It has been argued that ontologies could model information for social network offering a new set of opportunities.

Semantic Web techniques are gaining ground in structured areas such as scientific and technical contexts with rich data with intensive data processing and the willingness to reach a consensus about terms to create canonical specifications of vocabulary. In certain commercial applications, the potential profit from the use of well-structured and coordinated specifications of vocabulary will outweigh the sunk costs of developing or applying an ontology, and the marginal costs of maintenance. For instance, facilitating the matching of terms in a retailer's inventory with those of a purchasing agent will be advantageous to both sides. And the costs of developing ontologies may decrease as the user base of an ontology increases. If we assume that the costs of building ontologies are spread across user communities, the number of ontology engineers required increases as the log of the size of the user communities, and the amount of building time increases as the square of the number of engineers' efforts involved *per user* in building ontologies. Those for large communities gets very small and very quickly. Furthermore, as the use of ontologies spreads, techniques for their reuse, segmentation and merging will also become more familiar, and indeed there will be an increasing on reusing well-known base of ontologies.

Furthermore there is a perception of ontologies as top-down and somewhat authoritarian constructs, unrelated, or only tenuously related, to people's actual practice, to the variety of potential tasks in a domain, or to the operation of context. This perception may be related to the idea of the development of a single consistent Ontology of Everything, as the *OpenCyc* ontology. [6]

Such a wide-ranging and all-encompassing ontology may well have a number of interesting applications, but clearly it will not scale and its use cannot be enforced. If the Semantic Web is seen as requiring widespread buying to a particular point of view, then it is understandable that emergent structures like folksonomies begin to seem more attractive. This fits in general with calls for the dual and complementary development of Semantic Web technologies and technologies that exploit the self-organisation of the Web.

Folksonomies are building bottom-up classification systems, they are not a classification system. They are an unordered, flat set of keywords that are ranked by popularity. Ranking words by their popularity as in the tag-clouds can tell you a great deal about how groups of people are thinking and that information can be extremely useful, but it does not tell you much of anything about the relationships between words or concepts. In other words, there is not a system of rules.

Folksonomies cannot be compared with taxonomies, thesauri, or ontologies because they are not a classification system at all. They do not organize information, but aggregate individual acts of cataloguing ranked by popularity.

There are conceptual relationships between articles and web sites that are expressed when two or more are tagged with the same tag and these relationships can grow in very complex ways revealing a great deal about how people think and how some ideas can be related. But it's not clear how the overall set of conceptual relationships constitutes an organization of knowledge. This becomes particularly clear as the number of tags and sites multiply and the complexity of the tag and community relationships grows exponentially.

On the other hand, it is possible to browse through tags and citations and pick up a number of ideas of how other people have tagged a particular set of articles/web sites and thus be exposed to a variety of connections between concepts. Faceted representation, used to have semantic relations between folksonomies, can help searching between folksonomies

## 3. Joint meaning

Every kind of system based on a search engine tries to give the right answer to the user, but what is the meaning of the user query? And what the user have in mind?

In this work the act of search is considered as a communication act between the users of a community performed by a search engine with certain types of actions guided by the joint meaning here described.

According to the Speech Act Theory [7, 8] what a speaker wants to communicate depends on his/her intention being a function of that. The meaning of a linguistic expression in a language is determined by how it is used by a community of people. For a holistic view, the meaning involves connection between inferential and evidential connections and actions that people could take under various circumstances.

Referring to Herbet Clark, the meaning is jointly constructed by the *speakers* (users) and the *reader* (hearer) [9]. Community environment, like Facebook, are bringing the web on a more *Communication Acts* between a *speaker* and his/her *reader*.

The meaning of the communicative act, produced by the users of the community, appears to be collectively constructed by the *speakers* (as single user that contributes to his/her own community space) and by his/her *reader* called by some communities as "friend" (that can see the speaker community space).

We can consider $U = \{1, ..., n\}$ as a finite set of *n users*, supposing that $n–1$ users of $U$ communicate something to the *user u*. We can designate *u* as the *reader*, and the other $n–1$ users as the *speakers* $U–\{u\}$.

While speakers meaning is solely a function of the speakers' communicative intentions, joint meaning is a collective construal of the *reader u* and the *speakers* $U–\{u\}$. This work considers meaning as a conflict between reader meaning, understood as a personal mental state, and a collective construction between the personal mental as joint meaning. Every kind of meaning is understood as a different facet of knowledge, so making something common between the speakers and the reader means having a common faceted visualization and it means communicate. For the proposed work joint meaning can be considered as joint activities of two or more subjects' users that can develop together a *faceted interface* φ according to their communication. [10]

Communication can be defined by the fix point axiom of mutual belief that we can considered as the *fix point axiom of faceted interface for the communication*. The joint meaning of a communicative act from the *reader u* can coincide with one *speakers meaning* of the *speakers* $U–\{u\}$ in some cases. But in many cases joint meaning can be different from the reader meaning, whether or not reader's *u* communicative intention has been correctly understood by the *speakers* $U–\{u\}$ we have to deal with vagueness on ontological indeterminacy. The Semantic Web developers encodes all the information in an ontology filled with rules that say, essentially, that "Robert" and "Bob" are the same. But humans are constantly revising and extending their vocabularies, for instance at one times a tool might know that "Bob" is a nickname for "Robert," but it might not know that some people named "Robert" use "Rob", unless it is told explicitly.

Conversations could be seen as sequences of communicative acts, produced by two or more speakers, each of which has an associated speaker's meaning that depends on his/her communicative intentions.

*Joint meaning* is formed every time by a *reader u* and the *speakers* $U–\{u\}$ performing functions to maintain a shared view of what is said. It is not just common belief of what has been said, caused it may not coincide with the original speaker's meaning, but it is more a joint commitment of two or more subjects, who are obligated to each other to act coherently caring out deontic implications.

Joint meaning was performed between the *reader* and a frameworks of s*peakers* lattice. Every single speaker lattice was constituted by a formal context defined by an incidence of context. In the next section it will be shown how it was determined the incidence of context for a single speaker according to the Formal Concept Analysis. It is shown latter the joint meaning using the formal context extended to all the speakers.

### 3.1. Incidence of context of a speaker

Throughout this paper we will use the notion *concept* in the sense of *formal concept* as used in the ontological sense as well in Formal Concept Analysis (FCA), a branch of Applied Mathematics.

FCA is a method mainly used for the analysis of data, i.e. for investigating and processing explicitly given information. Such data are structured into units which are formal abstractions of concepts of human thought allowing meaningful comprehensible interpretation [11]. Central to FCA is the notion of a *formal context* that constitutes the dynamic corpus of folksonomies.

Considering facet representation and folksonomy tags: a triple $(f_n, t_n, i_n)$ was called *formal context, C*. We considered a *formal context* binding the use of RDF(S) constructs to remain in the first order logic.

Each triple (1) of the *formal context C* consists of a *facet* $(f_n)$ as subject, a *folksonomy tag* $(t_n)$ as predicate and an *incidence relation of context* $(i_n)$ representing an association, method invocation, or use-relationships.

$$(1) \qquad c_n = \left(f_n, t_n, i_n\right)$$

A set of such triples is called an RDF graph in which each triple is represented as a node-arc-node link.

Figure 1 depicts an example with:

*facet f* equal to Taiwan (the subject according to RDF);

*tag t* equal to Hot (the predicate according to RDF);

using *incidence i* (the object according to RDF) equal to Tropical, to disambiguate multiple matching.

Considering a *set of facets F*, a *set of folksonomy tags T* and a *set of incidence relations of context I*, a *set of formal contexts C* is defined by (2).

$$(2) \qquad C_n = \left(F_n, T_n, I_n\right)$$

Matching between facets $F$ and tags $T$ sets is defined by relation (3) allowing multiple associations among tags and faceted concepts.

$$(3) \qquad I \subseteq F \, x \, T$$

Multiple matching was disambiguated by updating a similarity degree associated for $\forall i \in I$

A triple $(F_n, T_n, I_n)$ is a formal concept of $(F, T, I)$ if and only if:

$$(4) \qquad F_n \subseteq F, \;\; T_n \subseteq T, \;\; T_n' = T_n \wedge F_n = F_n'$$

In other words, $(F_n, T_n, I_n)$ , is a *formal concept* if and only if the set of all attributes shared by the objects in $F_n$ is identical with $T_n$ and on the other hand $F_n$ is also the set of all the objects which have in common with each other the attributes in $T_n$. $F_n$ is then called the *extent* and $T_n$ the *intent* of the concept $(F_n, T_n, I_n)$.

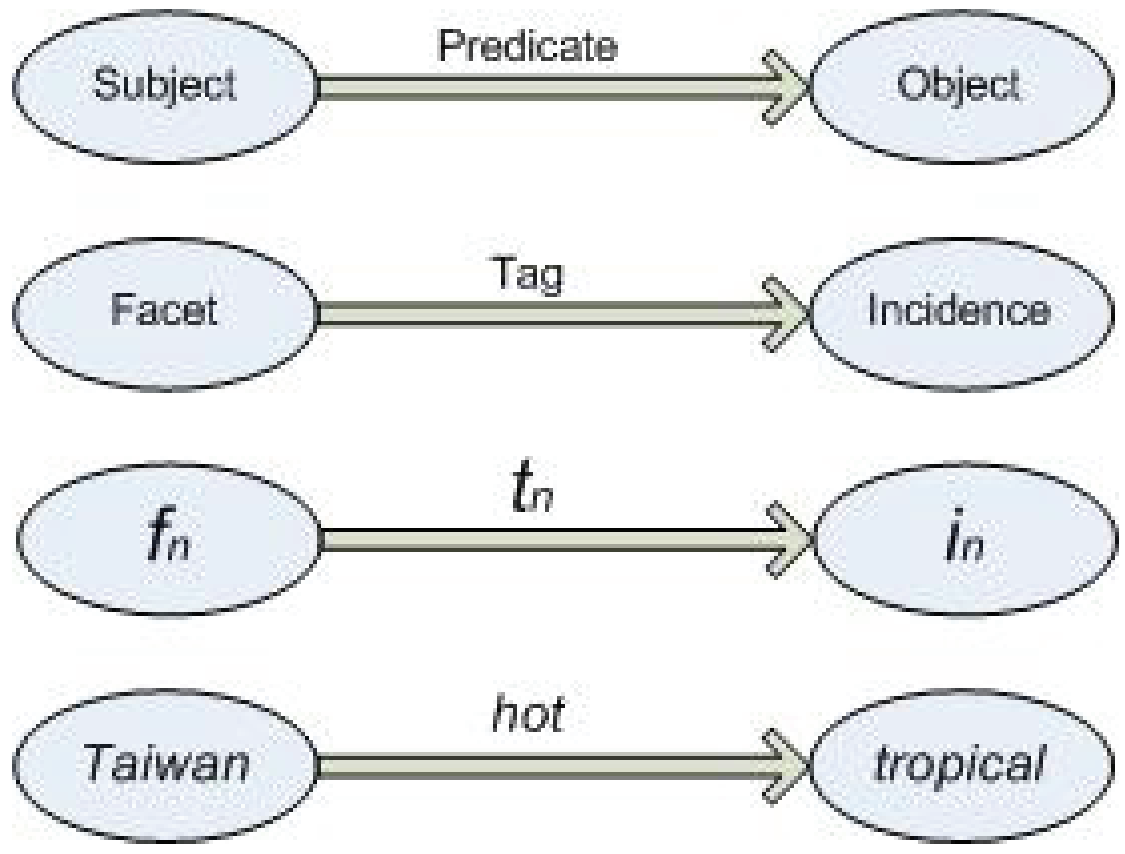

Fig. 1. *(f, t, i)* triple

The concepts of a given context are naturally ordered by the *subconcept-superconcept relation* [12] as defined by:

$$(5) \qquad \left(F_1, T_1, I_1\right) \leq \left(F_2, T_2, I_2\right) \Leftrightarrow F_1 \subseteq F_2$$
$$\left(\Leftrightarrow T_2 \subseteq T_1\right)$$
$$\left(\Leftrightarrow I_2 \subseteq I_1\right)$$

Thus, formal concepts are partially ordered with regard to inclusion of their extents or (which is equivalent) to inverse inclusion of their intent.

To obtain a tag hierarchy inside the set of folksonomy tags $T$ (i.e. a partial order) it was estimated how important every tag $t$ was for the concept topic represented by the facet $f$ in which $t$ is classified. To order the folksonomy tag element $t$ of the set of tags $T$ by the *subconcept-superconcept relation* we considered the DirectoryRank (DR) metric, introduced in [13], which organizes the tags $t$ that are assigned to the same facet $f$ based on the amount of information that every tag communicates about the facet $f$. DR defines the importance of a tag $t_i$ in a faced topic $f$ to be the sum of its topic relevance score V and its overall similarity to the other tag $t_n$ with which it correlates in the given facet $f$, as given by (6).

$$(6) \qquad DR_k\left(t_i\right) = V_k\left(t_i, t_n\right) + \frac{1}{n}\sum_{k=1}^{n} Sim\left(t_i, t_n\right)$$

For measuring the semantic similarity between tag $t$ it was used the Wu and Palmer metric [14] measuring the tags of the current contribution of the *speaker* with his previous one matching in the *formal context* deriving the similarity degree between pairs of tags $(t_1, t_2)$ as:

$$(7) \qquad Similarity\left(t_1, t_2\right) = \frac{2 * depth\left(LCS\left(t_1, t_2\right)\right)}{depth\left(t_1\right) + depth\left(t_2\right)}$$

Where *LCS* represents the *Least Common Subsumer* of the 2 tags in the Dynamic Corpus and *depth(t)* represents the length of the path from root of the *formal context* matching node.
A *root concept* was considered for any tag in the set *T* associated to faceted concept in *F* by means of *I*.
To determine the topic relevance score *V* of a folksonomy tag for each multiple matching between a facet and tags, $\forall i \in I$ , it was matched the co-occurrent tags *t* with the faceted concepts to construct a vector of connectivity degrees *i*. This vector *v[i]* was equal to the number of faceted concepts *f* associated to co-occurrent tags *t* and connected to the root concept *s* used by the single speaker. For the root tag concept *s* was considered the concept to be weighed by relevance. The corresponding sub-/superconcept partial order computed by FCA is depicted in form of a lattice in space. It was used the *reduced labelling* as described in [15] such that each faceted concept *f* and each tag *t* is entered only once in the lattice representation.
The matching with maximum similarity degree was selected.
The incidence *I* of the formal context can be represented in form of a matrix.
We accomplish this by creating for each node in the lattice (or grid) a faceted concept *f* labelled with the intent of the node as well as a subconcept of this concept for each element in the extent of that node representing a partial order.
Each faceted concept *f* is represented by a vector of folksonomy tags *t* chosen by the speaker.

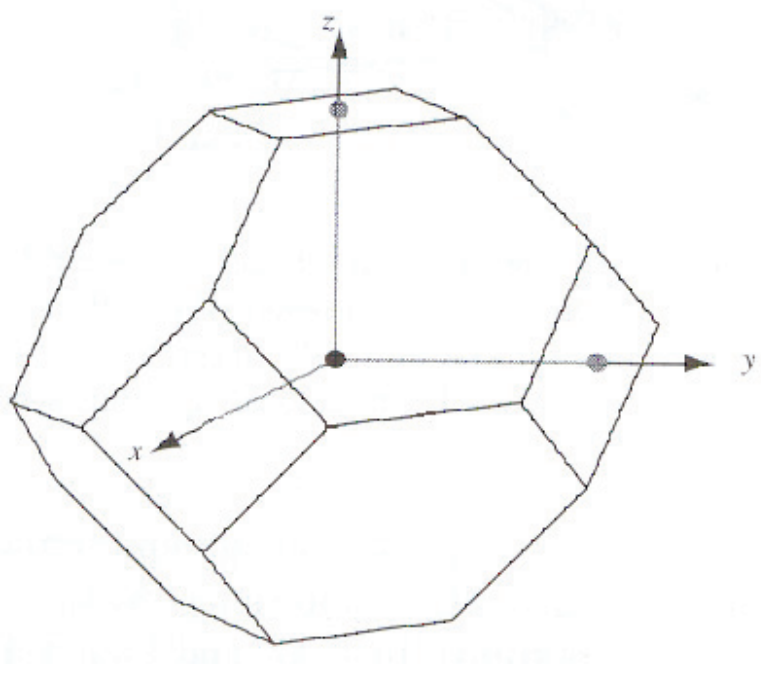

Fig. 2. A lattice composed by facets, tags and the relative incidence relations of context *(tn, fn, in)*

The values in the vector are the number of occurrences of each folksonomy tag in a facet. The linear tag vector is defined as the normalized inner product of the tag vectors. $t_1$, $t_2$ represent the two tag vectors.

$$V'(t_1, t_2) = \langle t_1, t_2 \rangle \quad (8)$$

$$V(t_1, t_2) = V'(t_1, t_2) \Big/ \sqrt{V'(t_1, t_1) V'(t_2, t_2)} \quad (9)$$

A labelled tags graph vector considers the whole network of tags relative to a facet *f*. In this vector design, a subject is represented as a vector network where each node (except the subject *S* to be weighed) is labelled with its tag classification category (see Figure 3).The similarity of two tags *t* is measured as seen above in (7).

Since it was aimed to compare folksonomy tags concept that are associated with tag networks rather than the inner structure of the networks, it was only compared the random walk paths starting from the subject S to be weighed. The random walks follow the tags directions from the root tag concept *s* to its directly *tags* used in the valuated subject *s* (see Figure 3). In each step, a random walk either jumps to one of the neighbours or stops by following a probability distribution. The longer a random walk path, the lower probability the path can exist.
Because nodes in the network are labelled by the folksonomy tags *t*, a random walk path is represented by a sequence of node labels (except the first one whose tag needs to be identified). By conducting pair wise comparisons of identical label sequence paths can be calculated. Such a probability is used as the tag value of a subject pair in the labelled tag graph:

$$V(L_1, L_2) = \sum_{s} \sum_{s'} v(s, s') P(s \mid L) P(s' \mid L') \quad (10)$$

Where $L_1$ and $L_2$ represent the formal context lattice associated with two subjects *s* and *s'* that are the random walk paths in the two grids. *P(s|L)* and *P(s'|L')* represent the probability random walk paths exist in networks. *v(s,s')* equals to *1* if the two sequences of labels of *s* and *s'* are identical. Otherwise *v(s, s')* equals *0*.
The labelled co-reference vector is a special case of a generic labelled graph vector which considers only single step random walk paths.

*1. S -> T10*
*2. S -> T4*
*3. S -> T5*
*4. S -> T4 -> T2*
...
*n. S -> T8 -> T11 -> T17*
...

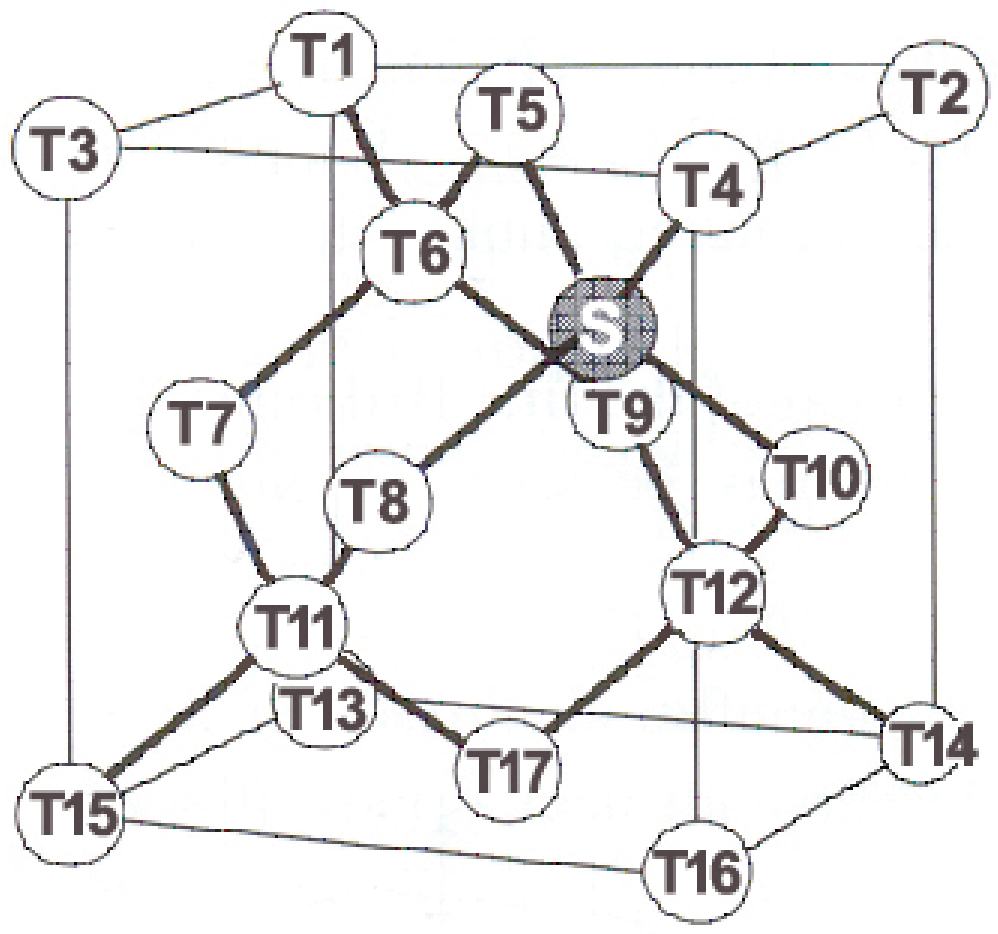

Fig. 3. Random walk paths among folksonomy tags in a formal context lattice

### *3.2. Joint meaning of concepts of all speakers*

A family of concepts that should be equivalent with each other was expressed by facet $f_n$ and the relative tag $t_n$ with the incidence $i_n$ matching from the folksonomies defined by the speakers *U–{u}*. This family of concepts was considered as a *formal context C* expressed by every single speaker.

The ordered set of all formal concepts of *(F, T, I)* was considered as the concept lattice.

For the *joint meaning* procedure was used the set of triple $(T_n, F_n, I_n)$ to represent a *formal context* as stated in (2) extended to all the formal contexts defined by all the speakers [16, 17].

$$(11) \qquad R = \left\{ C_o, \; ...., \; C_n \right\}$$

This family of all speakers *formal contexts* could be considered as a *framework R* related to a mathematical logic interpretation [18, 19].

$$(12) \qquad \sum\nolimits_n C_i \subseteq R$$

The multiple association *R* consists of a number of formal contexts that should be equivalent with each other expressed by facets $f_n$ and the relative tag $t_n$ with the relative incidence of concept $i_n$ matching the folksonomies defined by all the speakers toward every speakers formal context $C_n$.

The *joint meaning J*, expressed by (13) between the *reader u* and the *speakers U–{u}* towards the search engine, will be defined as a multiple association defined using a *Semantic Enrichment Method.* It consists of a number of a multiple association of formal contexts *R*, defined by the *speakers U–{u}*, and a domain ontology *O* used by the *reader u* that depend on the context of a user. As ontology *O* was used the translation of the Suggested Upper Merged Ontology (SUMO) into OWL adding concepts for domain specific content that wasn't well supported at the upper level. A matching between *O* and *R* is defined as a relation expressed by (13) allowing multiple associations among tags and concepts.

$$(13) \qquad J \subseteq R \times O$$

To disambiguate multiple matching was used an Hilbert space to evaluate compatible and incompatible Frameworks for faceted concepts defined by all the speakers contributors.

### *3.1. Quantum Frameworks*

The most important differences between the use of a predetermined ontology and folksonomies descriptions emerge when one considering several different multiple association of concepts *R* so different frameworks.

Using a predefined ontology, as long as the frameworks refer to the same ontology, there is no problem in combining the corresponding descriptions. But in the folksonomy case this is no longer true, and it is necessary to pay attention to the rules which state when descriptions can and cannot be combined.

To represent the multiple association *R* the notion of two and three-dimensional Euclidean space has been generalized to spaces with any finite or infinite number of dimensions. It was necessary to use a vector space endowed with an inner product and associated norm and metric. The preservation of some properties of Euclidean spaces in infinite dimensional function spaces is an Hilbert space.

Extending the Euclidean space based vector model - used in information retrieval - to Hilbert space, analogies were determined with the quantum logic. [21]

A Hilbert space is a set with a linear structure (space vector), which is called a scalar product, and that is guaranteed for completeness. An element on a physical state on the Hilbert space can be represented by a vector or by an appropriate linear combination of elements. In a Hilbert space the information of a quantum system may be determined by projecting the element state on an observable eigenstate. This operation generates a "dual" element which belongs to a new vector of Hilbert space, called the wave function.

The description of a quantum system requires the implicit adoption of some frameworks and compatible frameworks that are necessary for reasoning. The principles of quantum theory into present work are indebted to Omnes' ideas [20] for the quantum logic that defines the mathematical formalism of quantum mechanic. Quantum logic has formal similarities with Boolean algebraic structure, which provides the semantics of classical propositional logic.

Quantum logic was considered for its semantics representation, even if there is no commonly agreed syntax that could limits its applicability to specific problems.

A multiple association *R* defined in (12) is a finite collection of formal context $\{C_i\}$, for $i = 1, 2, . . . n$, that will be said to be mutually compatible if each formal context employs the same Hilbert space *H*, and if all the projectors associated with the different Boolean algebras $B_i$ commute with one another.

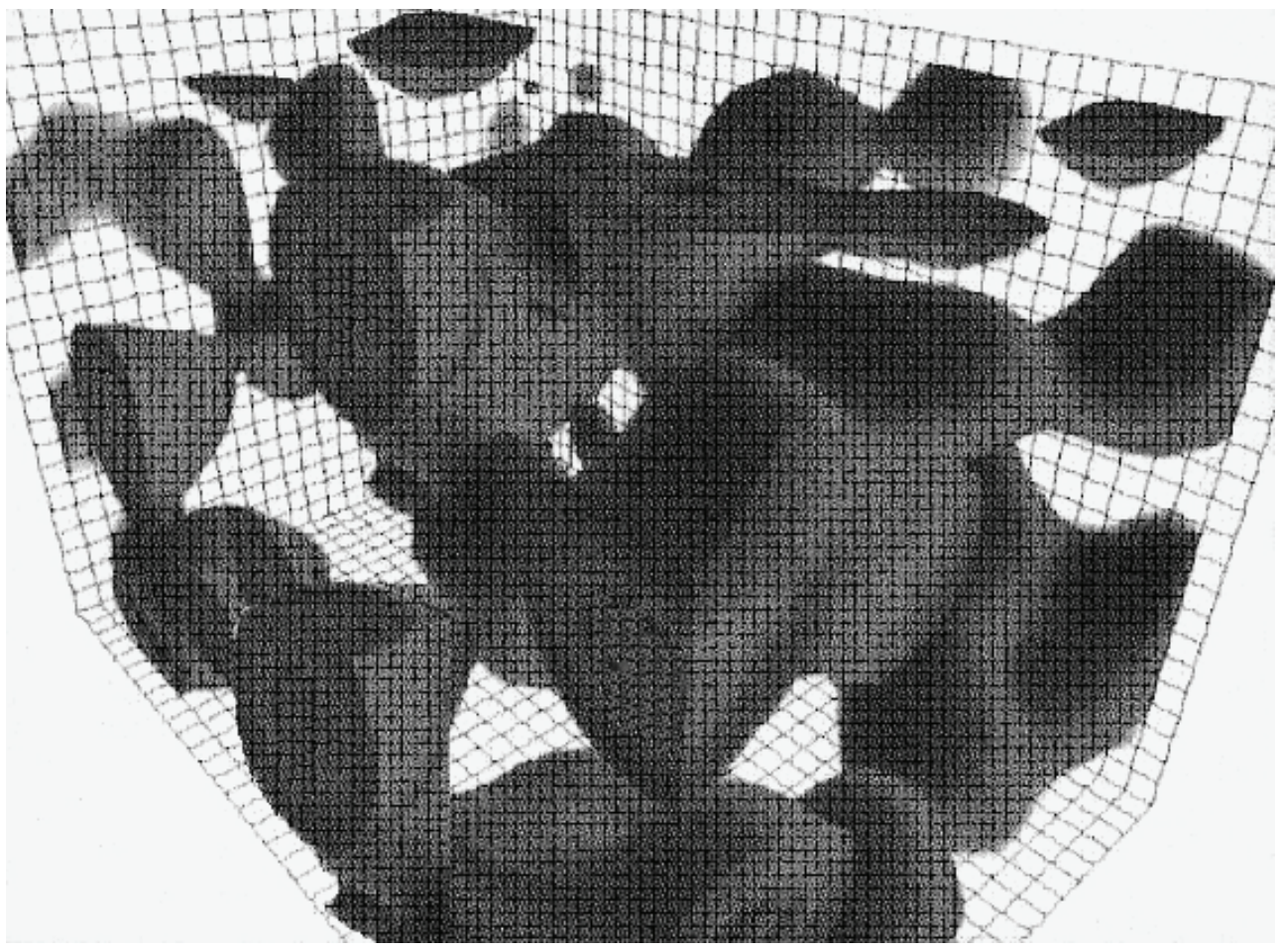

Fig. 4. A speaker node in the lattice with different faceted concepts.

Hilbert space $H$ is the counterpart for a quantum system of the classical phase space. The concept $(f_n, t_n, i_n)$, that describes a property of the system at a particular time, is associated with a closed subspace $P$ of $H$ defined by the orthogonal projection operator $P$ onto this subspace.
For example, if $p$ is the assertion that "a tropical temperature is between 5°C and 50°C", the subspace P is spanned by the eigenvectors of $H$ with eigenvalues which lie between 5 and 50 °C (assuming to be discrete). The assertion that the temperature lies outside this interval is defined by the negation $\neg p$ that corresponds to the *orthogonal complement of P* with projector $I-P$ (considering $I$ as the identity operator on $H$).

A formal context $C$ in the quantum logic is generated by a finite collection of concepts associated with projectors onto closed subspaces of $H$ provided these projectors commute with each other. The projector associated with a concept $p$ will be denoted by the spatial angle $\varphi(p)$. Using logical operations it is possible to obtain the concepts belonging to $R$ by the elementary concepts. The concepts can be defined by $\varphi$ onto projectors with the rules (14).

$$(14)\quad \varphi(\neg p) = I - \varphi(p), \quad \varphi(p \wedge q) = \varphi(p)\varphi(q)$$

Every logical operation can be built up using "not" and "and". [22]
A Boolean algebra is the smallest family $B$ containing the projectors associated with the concepts of $R$ where the operations of $\cap$ and $\cup$, acting on pairs of projectors, are defined by (15)

$$(15)\quad P \subseteq Q = PQ, \quad P \cup Q = P + Q - PQ$$

In $B$ the least elements is the zero operator 0 while the largest is the identity operator $I$. The framework $R$ can be defined as the collection of concepts generated from a set of elementary concepts by logical operations according to the Hilbert space $H$ and mapping concepts onto the Boolean algebra $B$ of commuting projectors as discussed above.
For notational consistency the concepts belonging to the different formal contexts should be mapped onto the projectors so that an elementary concept which occurs in more than one formal context is mapped to the same projector. Given a compatible collection $\{C_i\}$ there is a smallest formal context $C$ where concepts are generated from the union of the sets of elementary concepts for the individual $Ci$. Its Boolean algebra $B$ of projectors is the smallest one containing all the projectors of all the Boolean algebras $B_i$ associated with the different formal context $C_i$ in the collection.
Formal contexts not compatible between them are defined as incompatible. Formal contexts referring to the same folksonomy system in the same Hilbert space could be mutually incompatible because the projectors associated with one framework do not commute each other. This typical problem with quantum reasoning arose considering folksonomy since with ontology the counterparts of projectors always commute with one.

In a framework $R$ of the multiple associations of facet concepts any element of the quantum concept is mapped by the spatial angle $\varphi$. As for the classical descriptions, the truth of a quantum description is relative to a (implicitly) defined framework. For all statements provided by a framework R the assumptions of the argument, $a1, a2, \dots al$ are mapped onto a set of projectors $A1, A2, \dots An$ of $B$, whose product is shown in (16)

$$(16)\quad A = A1\, A2 \dots Al \quad ; \; Ai = \varphi\, (ai)$$

A statement $C$ for the framework $R$ is a valid conclusion if $AC = A$.

*3.1. Compatible and Incompatible Frameworks*

In the classical description there is no problem in combining frameworks of the multiple associations referring to the same ontology. While in the quantum case it is necessary to be careful to the rules that determine when descriptions can and cannot be combined cause every *speaker* contributor may use different folksonomy with own framework $R$ of multiple association.
Frameworks in a finite collection $\{R_i\}$, $i = 1, 2, \dots l$, are mutually compatible when each framework uses the same Hilbert space $H$, and all the projectors associated with the different Boolean algebras $B_i$ commute between them.
To have a notational consistency it is necessary that different frameworks statements are mapped with the same projector.
A smallest framework $R$, of the compatible collection $\{R_i\}$, contains all projectors. The union of the sets of elementary statements for the individual $R_i$ generates the statements of $R$. The smallest Boolean algebra $B$ of projectors contains all the projectors of all the Boolean algebras $B_i$ associated with the different frameworks in the collection. The smallest framework $R_i$ is generated by the compatible collection $\{R_i\}$.
Two or more frameworks which are not compatible are called incompatible. The distinctive problems associated with quantum reasoning arise from the existence of frameworks which use the same Hilbert space, and can thus (potentially) refer to the same physical system, but which are mutually incompatible because the projectors associated with one framework do not commute with those of another. There is nothing quite like this in classical mechanics, since the classical counterparts of projectors always commute with one another.
A significative quantum description must consist of a single framework and one of its statements, $(R, f)$, or a compatible collection of frameworks $\{R_i\}$ with associated facets $\{f_i\}$ defined by a collective description $\{(R_i, f_i)\}$.
A single master description $D$ can replace a collective description been the product of the projectors $F_i = \varphi(f_i)$ corresponding to the different facets as in (17), where $D$ is the framework generated by the collection $\{R_i\}$, and $d$ is any facet.

$$(17)\quad D = \varphi\,(d) = F_1\, F_2 \dots F_l$$

The rules for logical reasoning on quantum descriptions are similar to classical descriptions, but all the frameworks must be compatible.

A simultaneously true set of descriptions $\{(A_i, a_i)\}$, $i = 1, 2, \ldots n\}$, associated with a compatible collection of frameworks $\{A_i\}$,represents a set of assumptions for a logical argument.

If the union of the collections of frameworks $\{A_i\}$ and the collection of conclusion $\{Z_j\}$ is a compatible collection of frameworks the (18) can deduce a set of valid conclusion $\{(Z_j, z_j)\}$, $j = 1, 2, \ldots m$, assuming $A$ as the product of the projectors $Ai$ as seen on (17).

$$(18) \qquad \varphi\,(z_j)A = A$$

In this way is possible to deduce the original set of assumptions from a master description, so a master description $D$ acting as a single assumption can replace a set of assumptions $\{A_i\}$.

In this process of reasoning all the frameworks for the assumptions and conclusions must be compatible for the context considered.

Compatibility for quantum logic is not a transitive relationship: $A$ can be compatible with $Z$ and $Z$ with $W$ but at the same time $A$ can be incompatible with $W$.

The conclusion $(Z, z)$ can be deduced from an assumption $(A, a)$ if frameworks $A$ and $Z$ are compatible. After the conclusion $(W, w)$ can be deduced from the assumption $(Z, z)$ if frameworks $W$ and $Z$ are compatible. According to the classical logic it could be deduced that "if $a$ is true, then $w$ must be true". But according to the quantum logic this reasoning process is valid only if $A$ and $W$ are *compatible frameworks*.

Contradictions and paradoxes are due to inconsistent quantum reasoning that could come from not checking frameworks compatibility.

Different contexts may involve different and (possible) incompatible frameworks of multiple associations $R_i$ and they cannot be combined into a single facet. As an example, it is possible to have two valid contexts, based upon the same assumption $(A, a)$, one leading to the conclusion $(B, b)$ and the other to the conclusion $(C, c)$, where the frameworks $B$ and $C$ are incompatible. $c$ can be true relative to framework $C$, and $b$ can be true relative to framework $B$, but there is no framework true for both $b$ and $c$, it is not possible to check the validity of both.

To describe a quantum system it is necessary to choose a framework and there are many incompatible frameworks whose statements cannot be used in such description. But the choice of a framework does not influence the context being described.

For example considering a measurement, this requires the use of certain projectors for instance depending on the range and precision. According to compatibility it is not possible to consider something else represented by projectors which do not commute with the set of defined measures.

The type of exclusion which arises from incompatible frameworks is easily confused with, in fact it is quite different from the sort of exclusion which arises all the time.

In quantum logic if $p$ and $q$ are assertions represented by projectors $P$ and $Q$ which do not commute, they cannot be part of the same framework. While in classical logic descriptions when a property $p$ is true some other property $q$ must be false, because the corresponding subsets should not overlap they must be incompatible.

Therefore for a framework $P$ it is possible to consider the truth and falsity of $p$ but it has not meaning to consider the truth and falsity of $q$, since $q$ is defined for a framework Q.

According to quantum logic the truth of $p$ does not make $q$ false. Adopting a framework $P$ in which it makes sense to consider $p$, $q$ cannot be considered. Similarly, the combination "*p and q*", is not part of any framework, and therefore cannot be true or false, according to the quantum theory. In mathematical logic combinations of symbols must follow rules to form meaningful statements, for instance "*p and or q*" makes nonsense.

In the quantum logic, where the rules are different from classical logic, "*p and q*" ("*p or q*", etc.) are "nonsense" when $PQ \neq QP$

### *3.1. Quantum Collapse*

In a materialization of process of cognition in a closed system a *collapse*, or reduction into a classic situation, appears as process of evolution/action that end with the choice of one value. This value of the collapse will bring to the level of many values, that mean to choose one of the paths excluding other.

The collapse due to context may not always resolved by the context. [23] For instance consider "Fashion" in the context of "Style". We can easily assume there are at least two possible senses. One could deals with the women style, and the other deals with the men style. The distinction between measurements due to context needs to use human interaction with the *reader u* of the context. Threshold of the collapse is achieved by the appearance of a cognitive momentary event by the *reader u* that can reject the possibility of volition and return on his/her footsteps or try other alternative that creates a recognized pattern in brain memory due to past.

Joint meaning was so used to deal with a quantum collapse into a single state from a "quantum superposition" of states, a fundamental law of quantum mechanics to define the collection of all possible states that an object can have.

The interaction with the reader reduced to a single one state is represented by a facet shared with the *speakers* $U\text{–}\{u\}$.

In simplified terms, it is the condensation of different concepts, into a single occurrence, as seen by a *reader u.*

**4. Example of use**

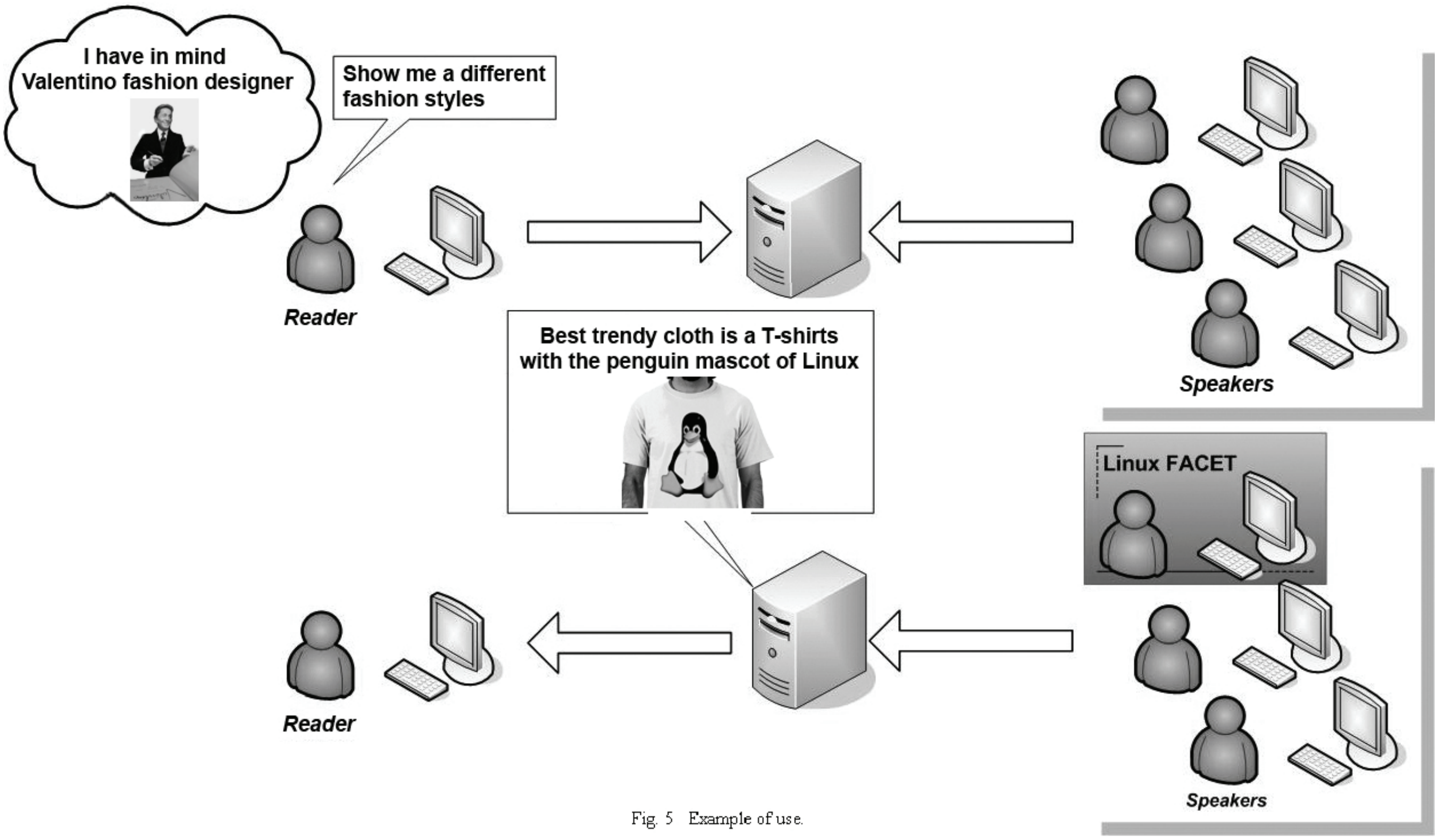

Fig. 5 Example of use.

In this section is discussed an experiment conducted to evaluate the effectiveness of the proposed work in improving the quality of retrieval results and therefore the users' web searches, of who was considered as the *reader* of the speakers contributions.

The use of joint meaning in faceted search for Semantic Web approach proposes to seamlessly integrate features of linguistic pragmatics and quantum logic to deal folksonomy tags. The novelty of this work lies in the capability of blending different paradigm and solutions to support innovative and effective dealing intrinsic ontological indeterminacy of folksonomy.

We can see some of the relevant features proposed with this work making reference to an example of use from a prototype of a people community website that has been developed locally for testing the proposed approach in a collaborative environment.

Suppose for example that David, our *reader u*, asks to a semantic rich engine (see Figure 5):

"I think I'm searching for style clothes" with the intention to search a fashion and glamour clothing having in mind Valentino fashion designer and imagine that the semantic search answers:

"The best trendy cloth is a T-shirts bearing Tux the penguin, mascot of Linux"

Clearly many fashions are popular in many cultures at any given time and may vary considerably within a society according to age, social class, generation, occupation, and geography as well as over time. The search engine has taken up David's statement as a search according to the actual trend between Linux supporters using the relative semantic relation between the asking for a style clothes, tag *T*, and the tag of the facet knowledge *F* of a Linux T-shirt expressed by *(F, T, I)* as a style clothes *R* as in (11).

This semantic relation was determined by a speaker, of the *speakers U–{u}* group, that considers a Linux T-shirt as a style clothes and it is expressed by the relation between the folksonomy tags and the facets towards an *incidence relation of context* $(f_n, t_n, i_n)$

David may now search answer, thus implying that his original statement was on searching fashion clothes, by asking for instance (see Figure 6):

"Pity. Well, show me a different fashion styles"

So he is refining his search on fashion facet of styles clothes.

At this point of the search the joint meaning of David's original statement is that it turned out as a search for a fashion style clothes. Then suppose that the semantic search engine answers:

"A T-shirt with written: I love fashion"

The search engine has taken up David's statement using the relative semantic relation between the fashion style, tag *T*, and the facet knowledge of who prefer the t-shirt, *F* expressed by *(F, T, I)* as a style clothes *I* as in (1).

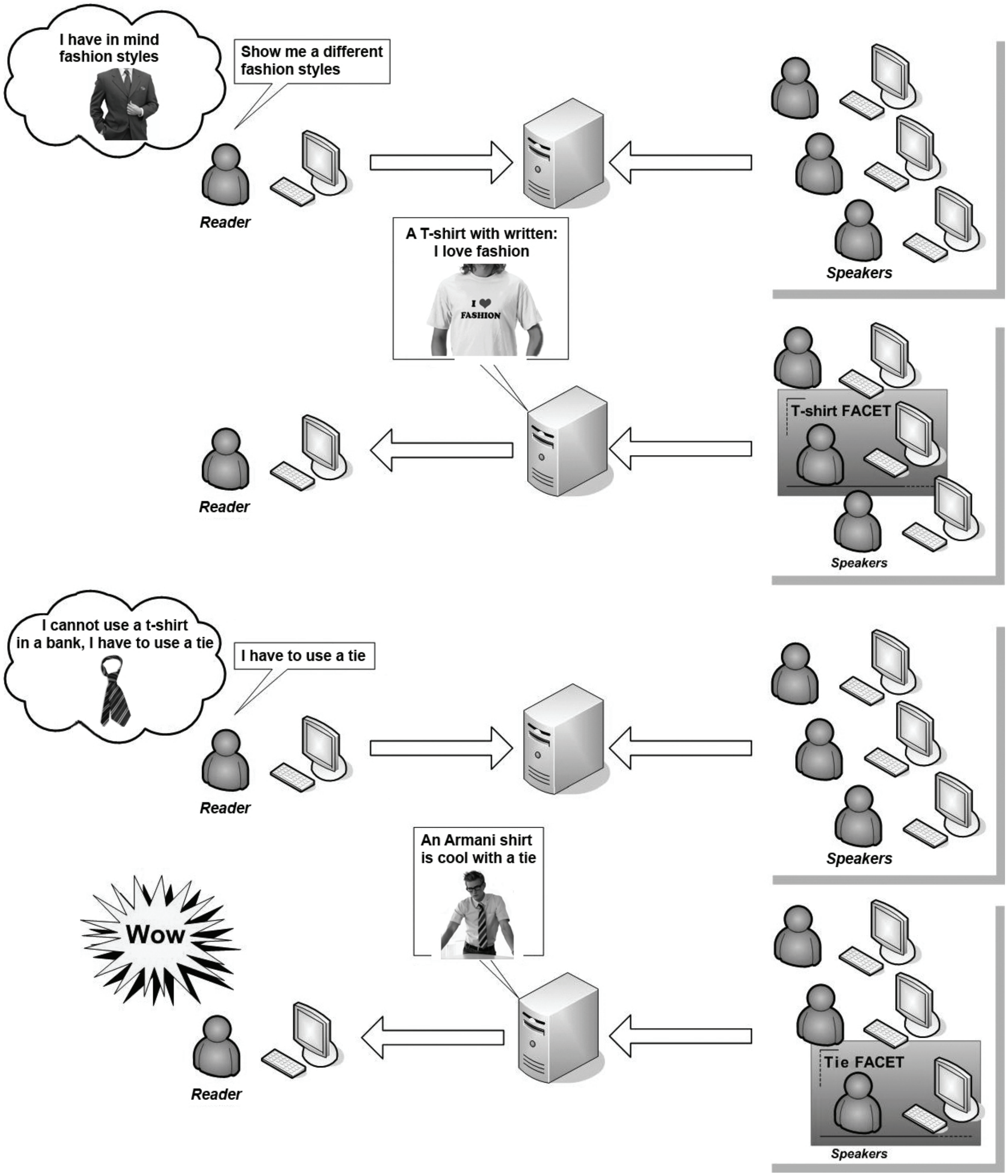

Fig. 6 Example of use.

But David dislike that and wants to use a tie at work, thus implying that his original statement was on searching fashion clothes for wearing a tie, by asking for instance:
"But I cannot use a t-shirt in a bank, I have to use a tie"

So the search engine looks for an opposite of t-shirt that could match a tie and answer with a shirt.

"An Armani shirt is cool at the moment with a tie" David initial meaning of his utterance was to search fashion clothing like the Valentino design, but now without violating a joint commitment he has accepted, as a matter of joint meaning, that such an utterance was a search on a glamour shirt to use with the popular mode tie for bank workers, and he is acting coherently to the search engine answers.

So he is taking the facet of the glamour shirt folksonomies from the fashion clothing knowledge according to the joint meaning using the semantic relation between glamour shirts and the popular mode tie for bank workers. This relation can change during the time cause fashion is relative to many variables. So fashion a fashionable Tux t-shirt for Linux community could be not so fashionable for fits in with the current popular mode of bank workers that use a tie.

Evolution of language that confounds parents who can't understand the slang of their teenagers also can trips up these systems. Folksonomies tags could be used to follow the evolution of the language to conceptualise different facets of a common knowledge to express users' preferences and needs since they allow users to add their own tags based on their interests.

Overwhelm criticism that folksonomy tags are ambiguous and uncontrolled terminology using more facets for a tag it will possible to reflect real users' views and their vocabulary.

## 5. Implementation

Created for people community the prototype website has been developed locally with the main functionalities for testing the proposed approach.

For the implementation was used the popular CMS Drupal with a plug-in that enables RDF and OWL output [23], and a themed AJAX interface is used to retrieve data integrating Flex. The website has been fully designed in its "traditional aspects" and a number of RIA features have been prototyped (to assess usability and effectiveness).

Faceted interface is defined in the Semantic Web by properties using triple to define the elements composing. Those are stored into the database access system, a SPARQL engines is integrated using RIA to combine the collaborative nature of Web2.0 with the ontologies of the Semantic Web.

Every faceted interface is composed by different kind of objects; those are identified by means of more than one aspect. So every object has more than one facet and the composition of every chosen facet compose the faceted interface between the *reader* and the *speakers* that can be identified by means of another facet.

The faceted interface has a *joint meaning* in multiple domains as a reference designation of the interface with respect to the *speaker* and the *reader* being related to one facet, see Figure 2.

A centralized management of the identification register is used for the objects. Using the Semantic Web the metadata information referred to any object can be arbitrarily voluminous and structured, having any desired information granularity. Being flexible it is not required the use of long identification. So the identification can easily be kept stable over time; while at the same time the content of the metadata can be adapted to current needs (e.g. restructured, increase of granularity).

The information could be fragmented, put into data bases, from which documents could be put together as needed including graphical presentations. [24]

Instead of "smart" economizing with computing power it had become more essential to describe things logically and straightforward in order to enhance functionality, exchangeability and communication. Another very important requirement had become emphasized, namely that the reference designations should be possible to use over the entire life cycle of the "objects".

The faceted interface is constructed using algorithm based on the joint meaning. Ranganathan's theory [11] could help us to automatically determine intuitive facets that belong to either of intuitive and unintuitive categories; ontologies contain the knowledge on the kind of facet.

Acquisition of faceted subject metadata is done by a *Folksonomy social tagging* used as a means towards building such structure consolidated towards the Semantic Web. Folksonomy are related to every kind of contribution like text, picture, audio, video or free code that are considered as a *portlet* framework's UI. These portlets are made either manually, writing code by the user, or via a portlet creation wizard. Each portlet is a component or service with its own folksonomy for the faceted taxonomy and faceted interface, been customizable onto one or more faceted view aggregation in the page, see Figure 3, according to the *joint meaning*.

### *5.1. Tags matching faceted concept*

It was considered two kinds of users: the *speakers group*, who participate actively, and the *reader*, who use lurking as participation.

The *speakers group* can: do/create, suggest and share documents.

The *reader* can browsing on documents searching items and reading text or looking for picture or video.

*Reader* performs search on the speakers contribution.

To achieve information the semantic search would be retrieve information according to the user knowledge on the topic.

The folksonomy is showing its weakness, one word may have several associative words. [25]

To understand the meaning of a word we have to consider the "meaning in use" as its use in the context of ordinary and concrete language behaviour which is sometimes referred to as a Wittgensteinian theory of meaning. [26] Several psychological experiments for human's word association were carried out on this. [27]

It would be so necessary allowing the conversion from the reader to the speakers contribution sharing a representation that could be used by both to define the context of the words. The solution is not to return to pure taxonomy but could be the use of folksonomy tags driven by their correlation with a faceted tags.

Although it may seem a restriction having a double contribution by the user, the system is not paradoxical, since it carries with it the user's discretion in deciding the focal points of the document to be classified together with a mechanism to co-ordinate in a more stringent rules thanks to a faceted system.
To order and retrieve those a semantic system for correlation and search will be important.
On the speaker interface a field was reserved to indicate the facet of the object considered and the corresponding field where contribute with the relative folksonomy tags according to the facet considered. (see Figure 7)
Insertion of facets and tags was favoured by the presence of suggested facets and relative tags according to the previous speakers. Those were showed according to the DirectoryRank (DR) metric

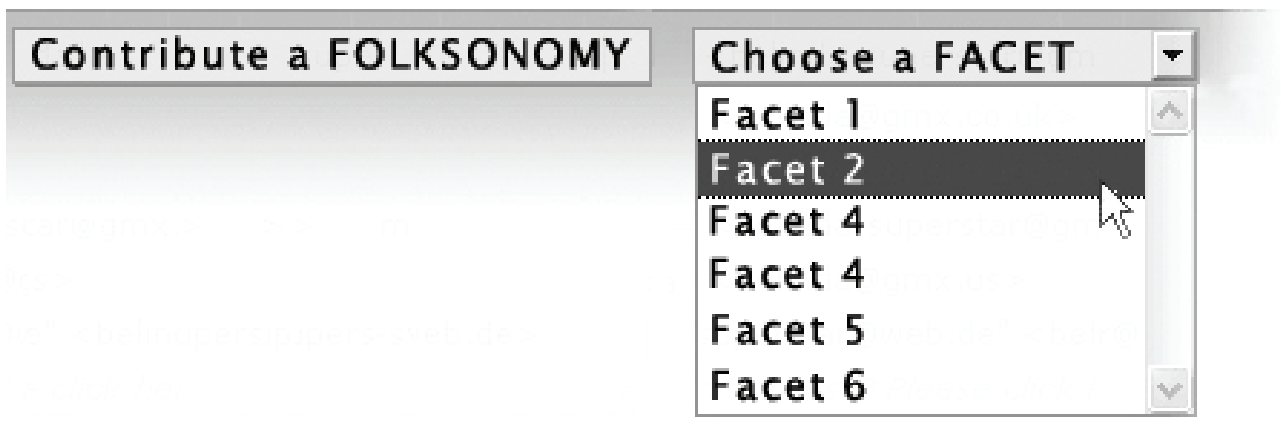

Fig. 7 Faceted and relative tags contribution by a speaker

It would be necessary to use data repositories to get the information from the participative systems.
Tags use different kinds of temporal entities as: instant, interval, duration, temporal unit

- instant: it is used for a specific begin and end;
- interval: between a specific begin and end;
- duration: between a not specific begin and end;
- temporal unit: defined as slot on the time or calendar.

Given a set of tags and a time structure the folksonomies can be considered as a set of version linearly ordered.

## 6. Related works

The use of "joint meaning" to design semantic search in a collaborative UI is an underexplored area that arose from the pragmatics [15, 16].
Different kinds of tools have been developed different aspects of the work. Three main classes are summarized below:

- Personalized Search

It aims at the retrieval of information that is tailored to the user interests. Search personalization has attracted a substantial amount of work over the last few years; most of witch addresses the challenge of user profiling. One approach to personalization is to have user explicitly describe their general search interests, which are stored in a personal profiles. Many commercial systems rely on personal profiles to personalize search results by mapping items to the same categories. For instance Google Personal asks users to build their search profiles by selecting topics of preferences. This profile can then be used to personalize retrieved results by mapping query-matching pages to the same topics. The work here proposed is different from the above approaches. The method aims at automatically capturing the user search preference (reader search) based on the analysis of the joint meaning according to the linguistic pragmatics

- Faceted Navigation

Most faceted navigation are used as interface for searching a large content database not considering different kinds of faceted visualization according to different kinds of reader (as done with "joint meaning") and using a fixed matching algorithm instead of a learning-based matching algorithm for automatic ranking of facet quality [28, 29]

- Social Networks

Nowadays the Semantic Web is looked by Social Networks for providing their bones. Adobe and Facebook cooperate to allow web developers to create RIA applications with the open source framework Flex. Facebook is working to use the Semantic Web on social network data used to predict some individual private trait [30].
Outside of Facebook, Microformats exist for tagging all kinds of information on ordinary web page. Microformats are simple standards that let you to give added significance to elements in HTML documents, and to expose information to third party software and services.
Social Networks provides features for listing the people you know, publishing contact information, and advertising planning events for group of friends and not dynamically for the *speaker* and the *single reader user*.

## 7. Evaluation

In this section is described the evaluation carried out studies in order to estimate the effectiveness of the proposed approach to match the preferences hidden behind the user queries, of the *reader u*, with the hidden knowledge expressed by the users contributions, the *speakers*. Finally it was experimental compared the performances of the proposed paradigm technique in delivering qualitative and user-relevant results to the expectation of the user (*readers*).
The evaluation was based on MiLE+ (Milano Lugano Evaluation Method) [31], which proposes an approach to usability evaluation under application-independent analysis (based on usability principles done by different experts) and application-dependent analysis (based on the requirements of the application, it is provided a step-by-step action guide for detecting the different problems with an assigned task, an example on Table I).

| Task: *Find cloth's men fashion to be used in a bank* | PREDICTABILITY | UNDERSTANDABILITY | RICHNESS | COMPREHENSIBILITY | GLOBAL SCORE FOR THIS TASK |
|---|---|---|---|---|---|
| Scores | 7 | 7 | 6 | 6 | 6.25 (average score) |
| Weights | 0.2 | 0.2 | 0.6 | 0.5 | |
| Weighted Scores | 0.9 | 0.9 | 2.3 | 1.6 | 6.2 (weighted score) |

*Table I Example of evaluation matrix.*
*Only readers interested in clothes men's fashion to be used in a bank can find the Armani shirt contributed by a speaker*

Application-dependent analysis focuses on the aspects of the user experience that can be assessed considering the actual domain of the application, the profiles of the intended users, the goals of the various stakeholders, or the context of use. At the heart of these analysis lays the concept of relevance, because relevant results are those which are interesting and useful to user. Among all was measured the precision and the recall of the answers of semantic search as stated in (19) and (20).

(19) $$Precision = \frac{\text{Number of relevant and retrieved}}{\text{Total numer retrieved}}$$

(20) $$Recall = \frac{\text{Number of relevant and retrieved}}{\text{Total numer relevant}}$$

*Precision* measures how well a system retrieves *only* the relevant documents.
*Recall* measures how well a system retrieves *all* the relevant documents. The relative importance of these metrics can vary based on the type of search and the search quality.
To evaluate how effective the proposed paradigm technique is in improving the search quality, it was relied on the following data: the queries that subjects issued, the set of the speakers information relevant to the reader queries, and the topic-importance values of the facets respect the topics in which they have been retrieved, as calculated by the DirectoryRank (DR) metric (6).
To collect these data was contacted 150 workers from an intranet community where the tool was tested for a period of one month. In particular were recorded all the queries the subjects issued during the experimental period and the search results that they chosen as responding their expectations. Ignoring queries with no results or where no results were considered it was collected a total of 8214 queries. Of those, 20% was overlapping queries (i.e. submitted more that once by the reader, and 12% were unique queries, submitted only once by the reader). On average, every participant issued 54.76 queries of which 12.87 were issued only once and 41.89 were submitted multiple times.

To identify the set of topics that are hidden behind the query traced it was carried out a user survey, where was asked to participants to keep a diary of their search during their participation in the evaluation, and indicate for every query, right after the query submission, the topic preference they had in mind. Table II summarizes some statistical on the experimental data.

| **Collection period** | **1 month** |
|---|---|
| Number of users | 150 |
| Number of queries | 8214 |
| Avg. # of queries/user | 54.76 |
| Avg. # of topic preference/user | 7.2 |
| Avg. # of visited speakers contribution/user | 8.5 |
| Avg. # of alternative words/user | 4.3 |
| Avg. # of nodes/incidence context | 4.7 |

*Table II Statistics on the experimental dataset*

The overall results confirm that people (on a sample of 150) prefer the faceted interface (83%) respect a classical search engine, finding it useful (95%) and easy-to-use (82%).

## 8. Overview and conclusion

Overall as Web dimension increases it's difficult for users to select queries expressing varying information needs in a distinguishable way by a search engine. Although refined queries may contribute to the improvement of retrieval results it is intrinsically limited by the preferences of the Web user with defined alternative wordings for expressing their search intentions. Since Web users are reluctant to provide explicit information on their personal preferences it would be necessary to learnt their preferences in order to disambiguate the current query and identify alternative query wordings that match both the initial query semantics and the user preferences.
This work aimed to explore a learning-based matching algorithm for automatic ranking of facet quality matching expressive queries and resource descriptions dealing with vagueness, incompleteness and inconsistencies in semantic on a "collective intelligence" (the ability of a group of people to solve problems than its individual members cannot represent).
Collective knowledge is composed by a fluid mix of experience, contextual information, values and expert insight that provides a framework for evaluating and incorporating new experiences and information. When we need such knowledge, we often cannot retrieve it because being tacit it is embedded in individual experience and involves intangible factors, such as personal beliefs, perspective, and the value system. Tacit knowledge underlies many competitive capabilities and is hard to articulate with formal language, it is unstructured or unorganized, and therefore remains hidden.
Human meanings are based on individual experiences, and logical axioms can only partially be reflected. An ontology is a formal specification of a conceptualization that can be understood as an abstract representation of the world or domain we want to model for a certain purpose. Ontologies are a critical aspect of the Semantic Web, being formal models of human domain knowledge they are difficult to build. Often there is no a discrete structure for single correct mapping on human knowledge.

A problematic complication trap behind ontologies which are nothing more than sets of logical axioms that can only partially represent the meanings of linguistic terms. Developing generic solutions that can find the hidden knowledge is extremely complex but this will be the biggest challenge for the developers of semantic technologies.

This work wanted to propose a direction for solutions able to make explicit and available the tacit knowledge hidden in the collective intelligence of a collaborative environment within organizations.

The environment was defined by folksonomies supported by common features on the faceted semantic search. For the representation and manipulation of folksonomies meaning this work propose the use of vector space models that can be extended to incorporate an analogy with the mathematical apparatus of quantum theory.

Information retrieval and quantum logic are based both on Hilbert space involving interaction between system and user (what was called as *reader*).Vector space retrieval has proven efficiency for information retrieval when there isn't a data behavioural to bear ranking algorithms involving a small number of types of elements and a few operations.

In this paper has been shown the use of the linguistic pragmatics in faceted search to deal with vagueness on ontological indeterminacy expressed by the user search (the *reader ontology O*) for personalized social search in a "collaborative" environment defined by folksonomies (multiple association of concepts *R* expressed by facets $f_n$ and the relative tag $t_n$ )

The main goal of this article has been to explore the basic argument on faceted search for folksonomies and useful aspects of pragmatics of dialogue towards the "joint meaning" understood as a joint construal of the creator of the community contents (*speakers*) and the user of the community contents (*reader*) thanks to the context adaptation using a faced taxonomy with the Semantic Web.

A prototype based on the proposed methodology was implemented to test its implementation by the actual technologies and its evaluation by a sample of users.

The ability to search on tags contributed by users with no a priori structured knowledge will be a crucial requirement of the information architectures that will make it possible to search in a personalized context-aware.

The described work has only scratched the surface of a huge problem being an initial step of a research program that will address several open issues:

1. a deeper comprehension of social commitment by subjects interaction as social reality intentionally constructed and how deontic affordances could be considered to produce joint meaning. [16] (see Carassa & Colombetti, 2009, for a first step in this direction);
2. enriching the methodological approaches by considering in depth different kind of RIA behaviours like: chat, multimedia synchronization, etc.;
3. developing automatic metrics for automatic facet ranking from the *Superconcept Formation System (SFS);*
4. working on the semantics of the conceptual model, to enable automated methodological approaches;
5. distilling comprehensive guidelines supporting the design RIAs;
6. measuring performance and optimizing the generated code;
7. continuing the industrial experimentation, by targeting other RIA platforms;
8. providing an UML profile and a visual notation for designing complex data-intensive Web application (WebML) for dealing with faceted interface using RIA for UI with the RUX-Method using the functionality of the existing web models. [32, 33]

## Acknowledgments

I would like to thank Professor Marco Colombetti for his advice on Knowledge Engineering. I am especially indebted to all the reviewers' detailed comments and constructive suggestions on the manuscript.

## References

[1] Zalka C. (1999). *Grover's quantum searching algorithm is optimal* (2746-2751), Physical Review A (Atomic, Molecular, and Optical Physics), Volume 60, Issue 4, October 1999

[2] Gruber T. (2007). *Ontology of Folksonomy: A Mash-up of Apples and Oranges* Published in Int'l Journal on Semantic Web & Information Systems, 3(2), 2007.

[3] Al-Khalifa H. S., Davis H. C. (2007). *Towards better understanding of folksonomic patterns* (163-166), HT'07 18th conference of Hypertext and hypermedia, ACM Press, New York, USA

[4] Angeletou S., Sabou M., Specia L. and Motta E. (2007). *Bridiging the gap between folksonomies and the semantic web: An experience report* (30-43), Bridging the Gap between Semantic Web and Web 2.0 (SemNet2007)

[5] Berners-Lee T. (2007) *The Future of the Web*, Testimony of Sir Timothy Berners-Lee CSAIL Decentralized Information Group Massachusetts Institute of Technology Before the United States House of Representatives Committee on Energy and Commerce Subcommittee on Telecommunications and the Internet.

[6] OpenCyc *ontology* http://www.cyc.com/

[7] Cooper A., Reimann R. (2007), *About Face 3 – The Essentials of Interaction Design* (156), Wiley.

[8] Pillan M., Sancassani S. (2004) *Il bit e la tartaruga* (122) Apogeo
[9] Sacco G.M., Tzitzikas Y. (2009). *Dynamic taxonomies aka Faceted Search*", Sprinter.
[10] Noruzi, A. (2004) *Application of Ranganathan's Laws to the Web,* Webology, Article 8
[11] Ganter B., Wille R. (1999). *Formal Concept Analysis. Mathematical Foundations*, Springer Verlag.
[12] Huang J., Huhns M. N. (2006). *Superconcept Formation System–An Ontology Matching Algorithm for Service Discovery.* WWW Workshop 2006, Beijing, PRC
[13] Krikos V., Stamou S., Kokosis P., Ntoulas A., and Christodoulakis D. (2005), *DirectoryRank: Ordering Pages in Web Directories* (17-22). Proceeding of 7th ACM Intl. Workshop on Web Information and Data Management 2005.
[14] Wu X., Palmer M. (1994) *Web semantics and lexical selection*. In 32nd ACL Meeting 1994.
[15] Herbert C. H. (1996) *Using Language* Cambridge University Press, Cambridge.
[16] Carassa A., Colombetti M. (2009). *Joint meaning*. Journal of Pragmatics, Volume 41, Issue 9, September 2009, Pages 1837-1854
[17] R. B. Griffiths (2002). *Nature and location of quantum information* Physical Review A (Atomic, Molecular, and Optical Physics), Volume 66, Issue 1, July 2002
[18] Manin Y. I. (1997). *A Course in Mathematical Logic* Springer-Verlag, New York
[19] Mendelson E. (1987) *Introduction to Mathematical Logic, 3d ed.* Wadsworth, Inc., Belmont, Calif.
[20] Omnes R (1994) *The Interpretation of Quantum Mechanics*, Princeton University Press, Princeton.
[21] Birkho G., von Neumann J. *The Logic of Quantum Mechanics (823)*, Ann. Math. 37
[22] Gudder S. (2007) *Handbook of Quantum Logic and Quantum Structures*. Elsevier, Amsterdam.
[23] Corlosquet S., Cyganiak R., Decke S., Polleres A. (2009) *Semantic Web publishing with Drupal.* DERI, Ireland.
[24] IEC 2009 *IEC 81346-1 Ed.1: Industrial systems, installations and equipmen,t and industrial products – Structuring principles and reference designations.*
[25] van Rijsbergen K. (2004). *The Geometry of Information Retrieval*. Cambridge.
[26] Wittgenstein L. (1953). *Philosophical Investigations.* Translated by G. E. M. Anscombe. Oxford: Blackwell.
[27] Karlgren J., Sahlgren M. (2001). *From words to understanding*. Foundations of Real-World Intelligence (294–308). Stanford: CSLI Publications.
[28] Bindelli S., Criscione C., Curino C. A., Drago M.L., Eynard D., Orsi G. (2008) *Improving Search and Navigation by Combining Ontologies and Social Tags* OTM Workshops 2008
[29] Diederich J., Balke W., and Thaden U (2007). *Demonstrating the Semantic GrowBag: Automatically Creating Topic Facets for Faceted DBLP* ACM IEEE Joint Conference on Digital Libraries 2007, Vancouver, Canada
[30] Lindamood J., Heatherly R., Kantarcioglu M., and Thuraisingham B. (2009). *Inferring Private Information Using Social Network Data*, WWWC 2009
[31] Bolchini D., Garzotto F. (2007). *Quality of Web Usability Evaluation Methods: An Empirical Study on MiLE+*; WISE 2007; Nancy, France
[32] Bozzon A., Comai S., Fraternali P., Toffetti Carughi G. *Conceptual modeling and code generation for rich internet applications (*353.360) ICWE 2006.
[33] Linaje M., Preciado J. C., Morales-Chaparr R., Sanchez-Figueroa F. (2008). *On the Implementation of Multiplatform RIA User Interface Components* (44-49), ICWE 2008.

**Massimiliano Dal Mas** is an engineer at the Web Services division of the Telecom Italia Group, Italy. His interests include: Ontology Engineering and Knowledge Base Systems, empirical computational linguistics and text data mining, user interfaces and visualization for information retrieval, automated Web interface evaluation. He received BA, MS degrees in Computer Science Engineering from the Politecnico di Milano, Italy. He won the thirteenth edition 2008 of the CEI Award for the best degree thesis with a dissertation on "Semantic technologies for industrial purposes" (Supervisor Prof. M. Colombetti).